\documentclass[reprint,prx,twocolumn,twoside,aps,groupaddress]{revtex4-2} 

\usepackage{amsmath,amssymb,amsfonts}
\usepackage{graphicx}
\usepackage{siunitx}
\usepackage{natbib}
\usepackage{braket}
\usepackage{multirow}
\usepackage[normalem]{ulem}
\usepackage{xspace}
\usepackage{xcolor}
\usepackage{url}
\usepackage[breaklinks]{hyperref}
\usepackage{braket}
\usepackage{lipsum}

\hypersetup{
    unicode=true,
    colorlinks=true,
    linkcolor=blue,
    citecolor=blue,
    urlcolor=blue
  }
\usepackage{breakurl}

\newcommand{\up}{$\ket{\uparrow}$\xspace}
\newcommand{\down}{$\ket{\downarrow}$\xspace}
\newcommand{\plus}{$\ket{\Psi^+}$\xspace}
\newcommand{\minus}{$\ket{\Psi^-}$\xspace}

\newcommand{\downdown}{$\ket{\downarrow\downarrow}$\xspace}
\newcommand{\downup}{$\ket{\downarrow\uparrow}$\xspace}
\newcommand{\updown}{$\ket{\uparrow\downarrow}$\xspace}
\newcommand{\upup}{$\ket{\uparrow\uparrow}$\xspace}

\newcommand{\pdown}{$P_\downarrow$\xspace}
\newcommand{\pdowndown}{$P_{\downarrow\downarrow}$\xspace}
\newcommand{\pupup}{$P_{\uparrow\uparrow}$\xspace}
\newcommand{\pdownup}{$P_{\downarrow\uparrow}$\xspace}
\newcommand{\pupdown}{$P_{\uparrow\downarrow}$\xspace}

\newcommand\transitionarrow{$\rightarrow$\xspace}
\newcommand\nottransitionarrow{$\not\rightarrow$\xspace}

\newcommand{\hzh}{$\mathrm{Hz}\times h$\xspace}
\newcommand{\kwcmsq}{$\mathrm{kW}/\mathrm{cm}^2$\xspace}

\usepackage{xr}
\begin{document}
\title{Long-lived entanglement of molecules in magic-wavelength optical tweezers}

\newcommand{\physics}{Department of Physics, Durham University, South Road, Durham, DH1 3LE, United Kingdom}
\newcommand{\jqc}{Joint Quantum Centre Durham-Newcastle, Durham University, South Road, Durham, DH1 3LE, United Kingdom}

\author{Daniel~K.~Ruttley}
\thanks{These authors contributed equally to this work.}
\affiliation{\physics}
\affiliation{\jqc}
\author{Tom~R.~Hepworth}
\thanks{These authors contributed equally to this work.}
\affiliation{\physics}
\affiliation{\jqc}
\author{Alexander~Guttridge}
\affiliation{\physics}
\affiliation{\jqc}
\author{Simon~L.~Cornish}
\email{s.l.cornish@durham.ac.uk}
\affiliation{\physics}
\affiliation{\jqc}

\begin{abstract}
Realising quantum control and entanglement of particles is crucial for advancing both quantum technologies and fundamental science. Significant developments in this domain have been achieved in a variety of systems 
\cite{Liu2019,Flamini2019,Kjaergaard2020,Adams2020,Yum2022}.
In this context, ultracold polar molecules offer new and unique opportunities due to their more complex internal structure associated with vibration and rotation, coupled to the existence of long-range interactions \cite{Cornish2024,DeMille2024}. However, the same properties make molecules highly sensitive to their environment \cite{Burchesky2021,Park2023,Langen2024}, impacting their coherence and utility in some applications. Here we show that by engineering an exceptionally controlled environment using rotationally-magic \cite{Guan2021,Gregory2024} optical tweezers, we can achieve long-lived entanglement between pairs of molecules using hertz-scale interactions. We demonstrate the highest reported fidelity to date for a two-molecule Bell state ($0.976^{+0.014}_{-0.016}$) and present the first realisation of a microwave-driven entangling gate between two molecules, preparing the molecules in a decoherence-free subspace. We show that the magic-wavelength trap preserves the entanglement, with no measurable decay over 0.5\,s, opening new avenues for quantum-enhanced metrology \cite{Zhang2023DecoherenceFree,DeMille2024}, ultracold chemistry \cite{Liu2024} and the use of rotational states for quantum simulation, quantum computation and as quantum memories. The extension of precise quantum control to complex molecular systems will allow their additional degrees of freedom to be exploited across many domains of quantum science \cite{Sundar2018,Sawant2020,Albert2020}.
\end{abstract}

\date{\today}

\maketitle

Precise control of quantum states and the generation of entanglement are essential for unlocking the potential of quantum systems for developing new technologies and exploring fundamental science.
Foundational work has focused on the quantum control a variety of systems, such as  trapped ions \cite{Yum2022}, superconducting circuits \cite{Kjaergaard2020}, neutral atoms excited to Rydberg states \cite{Adams2020}, quantum dots \cite{Liu2019}, and photons \cite{Flamini2019}, enabling many applications in quantum computing \cite{Cheng2023,Slussarenko2019,Flamini2019,Haeffner2008,Wendin2017,Krasnok2024,Bruzewicz2019,Saffman2016,Morgado2021}, metrology \cite{Giovannetti2004,Degen2017,Huang2024}, and simulation \cite{Altman2021}. 
Extending such control to more complex systems with more degrees of freedom, such as molecules, promises new advances in quantum metrology for fundamental physics \cite{Huang2024,DeMille2024}, the encoding of synthetic dimensions for quantum simulation \cite{Ozawa2019}, and high-dimensional quantum computing \cite{Wang2020Qudits,Sawant2020}. 

Ultracold polar molecules offer a rich internal structure associated with vibration and rotation, coupled to the existence of permanent electric dipole moments. These properties make molecules highly sensitive to a range of interesting phenomena \cite{Chin2009,Safronova2018,DeMille2024} and open up new prospects for studying ultracold chemistry \cite{Heazlewood2021,Liu2022}. In particular, the ladder of rotational states, with long radiative lifetimes, allow for storage of information and precise measurements over extended periods. Further, neighbouring rotational states are connected through electric-dipole transition moments, giving rise to long-range interactions that can be precisely controlled with external fields. These properties may be exploited for a wide range of applications \cite{Carr2009,Bohn2017,Softley2023}, including high-dimensional quantum computation \cite{Albert2020,Sawant2020,Cornish2024} and quantum simulation \cite{Wall2015,Sundar2018,Cornish2024}.

Recently, there has been rapid progress in the quantum control of molecules following the preparation of individual ultracold molecules in optical tweezers \cite{Liu2018,Anderegg2019,Cairncross2021,Zhang2022a,Ruttley2023,Vilas2024}.
Long-range interactions have been used to entangle pairs of molecules \cite{Holland2023Entanglement,Bao2023Entanglement,Picard2024Entangelement} and to interface them with other dipolar systems \cite{Guttridge2023}.
Protocols have been developed to simultaneously readout multiple molecular states and to realise global and local single-particle gates \cite{Ruttley2024,Picard2024SiteSelective}.
Further, mid-circuit detection and erasures of qubit errors have been demonstrated \cite{Holland2024Erasure}. However, despite recent advances, molecules prepared in rotational-state superpositions remain highly sensitive to their trapping environment. To sustain single-particle coherence for $\gtrsim100 \,$ms, rephasing pulse schemes are generally necessary \cite{Burchesky2021,Park2023,Langen2024}. This sensitivity restricts the interrogation time of individual molecules for precision metrology \cite{DeMille2024} and reduces the lifetime of the generated entanglement \cite{Holland2023Entanglement}, thereby limiting their effectiveness as long-lived quantum memories and sensors.

\begin{figure*}[t]
    \centering
    \includegraphics[width=\textwidth]{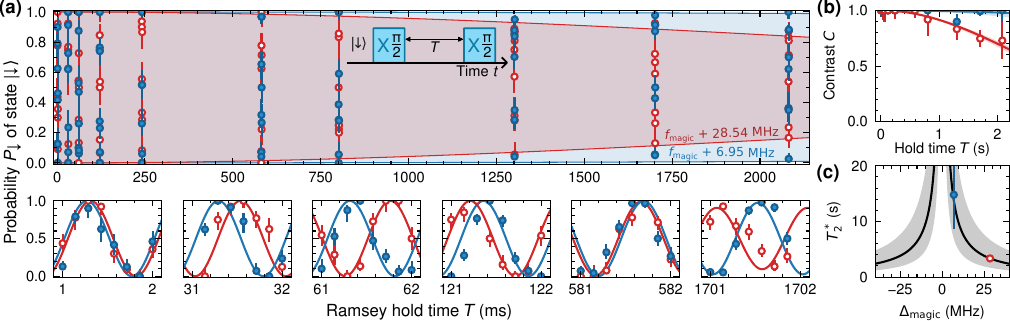}
    \caption{\textbf{Multi-second rotational coherence for individually trapped molecules.}  
    (a) Probability \pdown for a molecule to occupy the state \down after the Ramsey sequence (inset). The lower panels show detailed views of the upper panel. Blue filled (red empty) points correspond to molecules trapped in a tweezer with frequency $f_\mathrm{magic} + 6.95(7)$\,MHz ($f_\mathrm{magic} + 28.54(7)$\,MHz). (b) Ramsey fringe contrast as a function of the hold time $T$ between the Ramsey pulses. The solid lines are a fit to a Gaussian noise model. (c) The extracted $T_2^*$ times as a function of tweezer detuning $\Delta_\mathrm{magic}$ from $f_\mathrm{magic}$. The solid line represents the expected behaviour with 0.7\% intensity noise, while the shaded region shows the variation if this noise changes by a factor of two. Error bars in all plots show the $1\sigma$ confidence intervals.}
    \label{fig:ramsey-magic}
\end{figure*}

In this work, we create an exceptionally controlled environment for ultracold molecules by using magic-wavelength optical tweezers that eliminate single-particle decoherence on experimental timescales. 
This advance enables us to demonstrate entanglement of a pair of molecules with the highest reported fidelity to date, despite the hertz-scale spin-exchange interactions at the $2.8 \, \mu$m particle spacing we use. 
Additionally, we demonstrate for the first time the entanglement of two molecules using direct microwave excitation, opening up the prospect of using shaped pulses to engineer entangling operations robust to experimental imperfections. 
Both approaches result in long-lived entanglement owing to the use of a magic-wavelength trap, which will enable quantum-enhanced second-scale metrology, quantum simulation, and the encoding of quantum information within the rotational states of individually trapped molecules.

\subsection*{Magic-wavelength optical tweezers}

We begin by preparing molecules in a pristine environment that eliminates single-particle decoherence over typical experimental timescales.
We assemble individually trapped \textsuperscript{87}Rb\textsuperscript{133}Cs (hereafter RbCs) molecules in arrays of optical tweezers (Methods).
To engineer long-range interactions, it is necessary to drive rotational transitions that enable pairs of molecules to interact through dipolar spin-exchange interactions \cite{Wall2015}.
Generally, rotational decoherence arises primarily from differential ac Stark shifts that cause the energies of rotational transitions to fluctuate as molecules sample different trapping intensities \cite{Gregory2017,Langen2024}.

To eliminate these deleterious light shifts, we trap the molecules in optical tweezers formed from light at a magic wavelength in the vicinity of a weakly allowed electronic transition \cite{Bause2020,Guan2021} (Methods). 
This technique has not been previously used for individually trapped molecules, but has been used in bulk-gas samples \cite{christakis2023,Gregory2024} to achieve a rotational coherence time of 0.78(4)\,s without rephasing pulses \cite{Gregory2024}.
This method differs from earlier approaches for individually trapped molecules that used light at a magic polarisation \cite{Seesselberg2018Magic,Burchesky2021,Tobias2022,Park2023}.
For these experiments, the longest reported coherence time was 93(7)\,ms \cite{Burchesky2021}, limited by second-order couplings between hyperfine states \cite{Blackmore2020,Langen2024}. 
By using magic-wavelength light, we eliminate these couplings to first- \textit{and} second-order \cite{Blackmore2020}.

We determine the magic wavelength through detailed molecular spectroscopy.
All molecules in our experiment begin in the rovibrational ground state 
\down which we couple to the rotationally excited state \up using microwave radiation (Methods).
We measure the differential ac Stark shift $h\Delta\alpha_\mathrm{ac}$ of the transition $\ket{\downarrow}\rightarrow\ket{\uparrow}$ with a Ramsey procedure \cite{Hepworth2024}.
When the tweezers are formed from light at the magic frequency $f_\mathrm{magic}$, we eliminate $\Delta\alpha_\mathrm{ac}$.

We probe the rotational coherence of the molecules using the Ramsey interferometry sequence illustrated in the inset of Fig.~\ref{fig:ramsey-magic}(a).
We apply two $\pi/2$ pulses with a hold time $T$ between them. Both pulses have the same phase which we use to define the $\hat{x}$ axis of the Bloch sphere. The microwaves drive the transition \down\transitionarrow\up with Rabi frequency $\Omega=5.0(1)$\,kHz at a detuning $\Delta\approx1$\,kHz.
The first pulse prepares each molecule in the state $(\ket{\downarrow}+i\ket{\uparrow})/\sqrt{2}$.
Phase accumulates between \down and \up during the hold, and the second pulse projects this onto the states \down and \up.
The populations of these states oscillate as a function of $T$ with frequency $\nu = \Delta-\Delta\alpha_\mathrm{ac}$.
We note that this sequence does not include any rephasing pulses.

Figure~\ref{fig:ramsey-magic}(a) shows the probability \pdown of molecules occupying the state \down as $T$ is varied.
We use a multistate readout scheme to measure the internal state of each molecule \cite{Ruttley2024} and correct for state-preparation and molecule-loss errors with postselection (Methods).
For this measurement, we prepare a two-molecule array with separation $8.6(2)\,\mu$m for which we expect an interaction strength $\sim 0.2$\,\hzh.
For this reason, we use $T\lesssim2\,$s so that interactions can be neglected.
In order to measure some decoherence over this timescale, we detune the first tweezer (blue filled points) from $f_\mathrm{magic}$ by $6.95(7)$\,MHz and the second tweezer (red empty points) by $28.54(7)$\,MHz.
The data are fitted with a damped sinusoidal functions, and the difference in the measured transition frequencies is within 10\% of the expected value (Methods).

The decoherence measured in Fig.~\ref{fig:ramsey-magic}(a) indicates that the dominant dephasing mechanism is shot-to-shot noise in the tweezer intensities. 
Fig.~\ref{fig:ramsey-magic}(b) shows the Ramsey contrast $C$ as a function of $T$, obtained by fitting each cluster of points in Fig.~\ref{fig:ramsey-magic}(a) independently.
We fit $C$ with a Gaussian noise model $C(T) = e^{-(T/T_2^*)^2}$ from which we extract the Ramsey coherence times $T_2^*$ as 15(6)\,s and 3.3(2)\,s for the two tweezers respectively. 
Fig.~\ref{fig:ramsey-magic}(c) shows these $T_2^*$ times as a function of the detuning $\Delta_\mathrm{magic}$ from $f_\mathrm{magic}$. 
We model the expected behaviour assuming Gaussian noise in tweezer intensities (Methods).
The data are consistent with intensity noise (standard deviation) of 0.7\% (solid line). 
The shaded region shows the expected behaviour if this noise varies by a factor of two. 
This level of intensity noise is consistent with ex-situ measurements of the tweezer powers.
With this level of intensity noise, our model predicts that, for $|\Delta_\mathrm{magic}| \lesssim 0.5\,$MHz, $T_2^*$ exceeds a few minutes.
Therefore, we have effectively eliminated rotational decoherence due to the trapping potential on timescales relevant to our experiment.

\begin{figure*}[t]
    \centering
    \includegraphics[width=\textwidth]{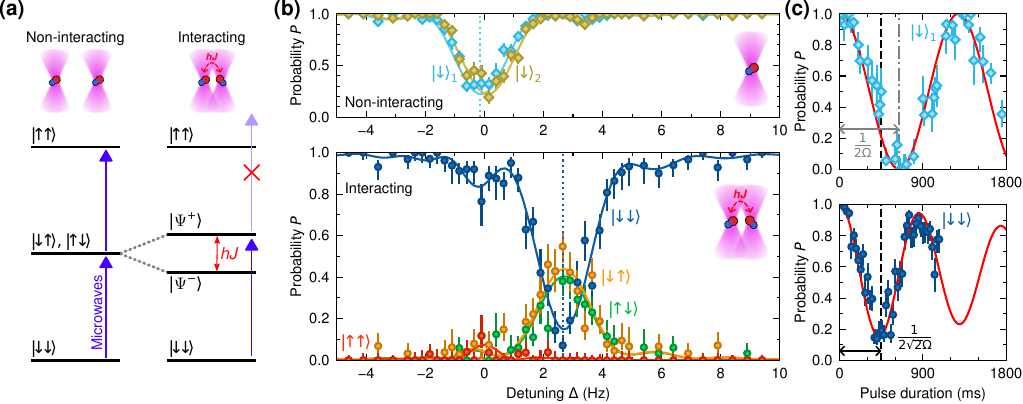}
    \caption{\textbf{Microwave spectroscopy of a pair of interacting molecules.} (a) Eigenstates of the Hamiltonian $H$ in the non-interacting (left) and interacting (right) cases. Interactions cause the single-excitation states \downup and \updown to couple to form two entangled states $\ket{\Psi^\pm} \equiv (\ket{\downarrow\uparrow}\pm\ket{\uparrow\downarrow})/\sqrt{2}$ which have an energy difference of $hJ$. We drive transitions between the eigenstates with microwaves. (b) Microwave spectroscopy of single molecules (upper panel) and pairs of molecules (lower panel) using a square spectroscopy pulse of duration 441\,ms and detuning $\Delta$ from the mean frequency of the single-molecule transitions. We show the probability of occupying different states, presented using different colours (see text). (c) Enhancement of the Rabi frequency by $\sqrt{2}$ when driving the two-molecule transition \downdown\transitionarrow\plus (lower panel) compared to the single-molecule transition \down\transitionarrow\up (upper panel). The data in all panels are fitted simultaneously with a single set of free parameters; the solid lines show these fits. Error bars show the $1\sigma$ confidence intervals.}
    \label{fig:mw-spec}
\end{figure*}

\subsection*{Rabi spectroscopy of interacting molecules}
We use precision microwave spectroscopy to investigate dipolar interactions between pairs of molecules.
The Hamiltonian that describes a pair of molecules interacting via the dipolar spin-exchange interaction in the presence of microwave coupling between $\ket{\downarrow}$ and $\ket{\uparrow}$ with a Rabi frequency of $\Omega$ is  \cite{Wall2015}
\begin{equation}
H = H_\mathrm{mol}^{(1)} \otimes\mathcal{I}^{(2)} + \mathcal{I}^{(1)}\otimes H_\mathrm{mol}^{(2)} + H_\mathrm{int}\,,
\label{eq:ham}
\end{equation}
where  $H_\mathrm{mol}^{(i)} = \frac{1}{2}h\Omega\left(\sigma_i^+ + \sigma_i^-\right) - h\Delta_i\ket{\uparrow}_i\bra{\uparrow}_i$ is the single-particle Hamiltonian of molecule $i$ and $\mathcal{I}^{(i)}$ is its identity operator.
$H_\mathrm{int} = \frac{1}{2}hJ\left(\sigma_{1}^+\sigma_{2}^- + \sigma_{2}^+\sigma_{1}^-\right)\,$ is the interaction Hamiltonian and $\sigma^+_i\equiv\ket{\uparrow}_i\bra{\downarrow}_{i}$ ($\sigma^-_i\equiv\ket{\downarrow}_i\bra{\uparrow}_{i}$) is the raising (lowering) operator for molecule $i$.
$hJ$ is the interaction strength and $\Delta_i$ is the microwave detuning from the transition $\ket{\downarrow}_i\rightarrow\ket{\uparrow}_i$.
We allow for the fact that there may be a small difference $\delta \equiv \Delta\alpha^{(2)}_\mathrm{ac}-\Delta\alpha^{(1)}_\mathrm{ac} = \Delta_1 - \Delta_2$ in the differential ac Stark shifts of the molecules as they are in different traps and denote pair states as $\ket{ab}\equiv\ket{a}_1\otimes\ket{b}_2$.

Figure \ref{fig:mw-spec}(a) shows the eigenstates of $H$.
The left panel shows the non-interacting limit ($J\rightarrow0)$ when the two tweezers are perfectly magic ($\Delta\alpha^{(1)}_\mathrm{ac},\Delta\alpha^{(2)}_\mathrm{ac}\rightarrow0)$.
The microwaves couple the ground state \downdown to the degenerate states \downup and \updown in which there is a single rotational excitation.
These states are coupled to the doubly excited state \upup.
When the interaction between the molecules becomes significant ($J\gg\delta$), the singly excited states become coupled.
The right panel of Fig.\,\ref{fig:mw-spec}(a) shows the resultant eigenstates, which include the two entangled states $\ket{\Psi^\pm} \equiv (\ket{\downarrow\uparrow}\pm\ket{\uparrow\downarrow})/\sqrt{2}$.
The energy difference between these states is $hJ$.
Microwaves can couple between the symmetric states of the triplet manifold $\{$\downdown, \plus, $\ket{\uparrow\uparrow}\}$ such that the transition \downdown\transitionarrow\plus is allowed. In contrast, the antisymmetric singlet state \minus is decoupled \cite{Turchette1998,Hughes2020}.

We directly probe these energy levels with microwave spectroscopy.
We form two near-magic tweezers that are separated by $2.78(5)\,\mu$m and use a square spectroscopy pulse of duration $441$\,ms with $\Omega = 780(7)$\,mHz.
We study the non-interacting case by postselecting on experimental runs in which only a single molecule was formed (Methods).
The upper panel of Fig.\,\ref{fig:mw-spec}(b) shows the spectroscopy in this case.
The blue (gold) points show the probability \pdown that the molecule remains in the state \down after the spectroscopy pulse when it is in the first (second) tweezer.
The microwave detuning $\Delta \equiv (\Delta_1+\Delta_2)/2$ is relative to the mean frequency of the single-molecule transitions which differ by $\delta=220(40)$\,mHz.

When two molecules are present, we can directly excite to the state \plus when  $\Delta \approx J/2$.
~This can be seen in the lower panel of Fig.~\ref{fig:mw-spec}(b) where data are from experimental runs where a molecule was prepared in each tweezer.
Excitation out of the state \downdown (blue) at zero detuning is suppressed due to the interaction shift.
Similarly, two-photon excitation to the state \upup (red) is prevented via a rotational blockade effect \cite{Chae2021} as illustrated in Fig.~\ref{fig:mw-spec}(a).
At $\Delta \approx J/2$, the slight preferential occupation of the state \downup (orange) over the state \updown (green) is due to the positive value of $\delta$ (Methods).

We verify that we drive a collective excitation by measuring an enhancement of the Rabi frequency for the transition \downdown\transitionarrow\plus compared to single-molecule transition.
We use the same experimental routine as above, but vary the duration of the square spectroscopy pulse.
The upper panel in Fig.~\ref{fig:mw-spec}(c) shows \pdown when a molecule is prepared in the first tweezer and the second tweezer is empty.
For this measurement, we set the microwave detuning to be resonant with the transition $\ket{\downarrow}_1\rightarrow\ket{\uparrow}_1$ [Fig.~\ref{fig:mw-spec}(b), light dotted line].
As expected, we see oscillations at Rabi frequency $\Omega$ with a $\pi$-pulse duration of $\sim 640$\,ms.
The lower panel in Fig.~\ref{fig:mw-spec}(c) shows \pdowndown when two molecules are prepared in \downdown and $\Delta = 2.58(1)$\,Hz [Fig.~\ref{fig:mw-spec}(b), dark dotted line].
We drive the transition \downdown\transitionarrow\plus with the enhanced Rabi frequency $\sqrt{2}\Omega$ that we expect for a collective excitation of two particles \cite{Saffman2002,Gaetan2009} and a $\pi$ pulse takes $\sim450$\,ms.

The dynamics of our system are well described by the Hamiltonian in Eq.~\eqref{eq:ham} which allows us to fit the interaction strength $J$.
Our molecules are predominantly, but not exclusively, formed in the three-dimensional motional ground state \cite{Ruttley2024}.
This causes shot-to-shot noise in $J$ as the separation averaged over the molecular wavefunctions varies.
We incorporate this in our model with a Monte Carlo method: the dynamics are averaged over 200 iterations where $J$ is sampled from a Gaussian distribution with mean $\langle J\rangle$ and standard deviation $\sigma_J$.
We perform a least-squares fit to obtain $\langle J\rangle = 5.20(5)$\,Hz and $\sigma_J = 1.0(1)$\,Hz. 
This is consistent with expected interaction strength in our system (Methods) and the solid lines in Fig.~\ref{fig:mw-spec} show the dynamics predicted by this model.

\subsection*{Spin-exchange entanglement}
As a benchmark for the exceptional control that we realise in our experiment, we turn our focus to entangling pairs of molecules with the observed hertz-scale dipolar interactions reported above.

\begin{figure*}[t]
    \centering
    \includegraphics[width=\textwidth]{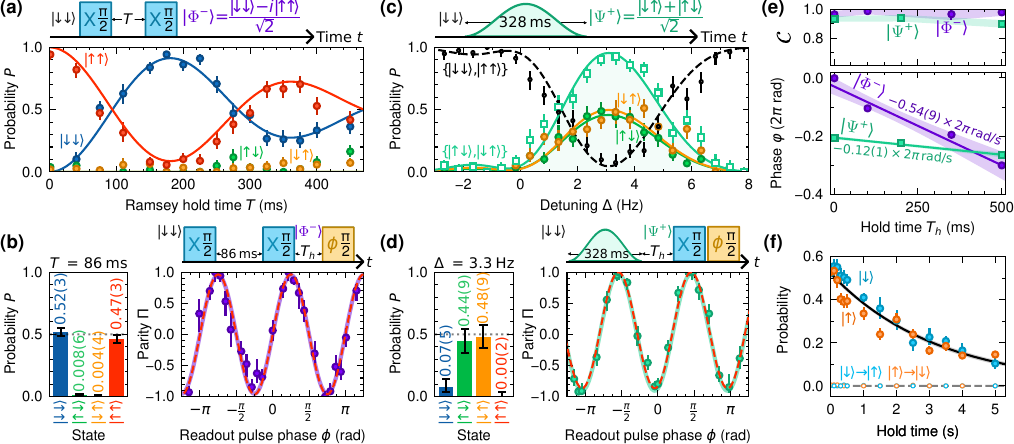}
    \caption{\textbf{Preparation and characterisation of long-lived molecular entangled states.} (a) Molecule entanglement using spin exchange. We show the state probabilities $P$ as a function of $T$ in the Ramsey sequence shown. The colours are as in Fig.~\ref{fig:mw-spec}. (b) Measurement of the fidelity $\mathcal{F}$ with which we entangle molecules using $T=86$\,ms. Left: State populations after the Ramsey sequence. Right: Parity $\Pi$ measured as a function of the phase of the readout pulse (see text) with a fit (dashed red line) and a model prediction (solid purple line, Methods). From these data, we extract $\mathcal{F} = 0.976^{+0.014}_{-0.016}$. (c) Molecule entanglement with direct microwave excitation. We show the state probabilities $P_{\downarrow\downarrow}+P_{\uparrow\uparrow}$ (black), $P_{\downarrow\uparrow}+P_{\uparrow\downarrow}$ (empty green), \pupdown (filled green), and \pdownup (orange) as a function of detuning $\Delta$. (d) Measurement of the fidelity $\mathcal{F}$ with which we entangle molecules with $\Delta=3.3$\,Hz. Left: State populations after the microwave pulse. Right: Measurement of the parity $\Pi$ with a fit (dashed red line) and a model prediction (solid green line). From these data, we extract $\mathcal{F} = 0.93^{+0.03}_{-0.05}$. (e) Long-lived entanglement for molecules in $\ket{\Phi^-}$ (purple) and \plus (green). Top: Entanglement coherence $\mathcal{C}$ after holding the entangled state for a time $T_h$. The shaded regions are a guide to the eye. Bottom: Phase $\varphi$ of parity oscillations as a function of $T_h$. The solid lines show linear fits and the shaded regions show the $1\sigma$ uncertainties of the fits. (f) Lifetime of single molecules in \down (blue) and \up (orange). We do not observe significant bit-flip errors (empty points). Error bars in all plots show the $1\sigma$ confidence intervals.}
    \label{fig:entanglement}
\end{figure*}

First, we entangle a pair of molecules using resonant energy exchange.
This method has recently been used to entangle pairs of CaF \cite{Holland2023Entanglement,Bao2023Entanglement} and NaCs \cite{Picard2024Entangelement} molecules with interactions orders of magnitudes stronger than those in our system. 
To generate this entanglement, we prepare the molecules in rotational superpositions and wait to allow resonant exchange of energy between the pair.
We use the Ramsey pulse scheme shown in Fig.~\ref{fig:entanglement}(a): we apply two $\pi/2$ pulses ($\Omega = 5.0(1)$\,kHz) on the transition \down\transitionarrow\up to a pair of molecules in \downdown with a hold time $T$ between the pulses. 
Ideally, the effect of this pulse sequence is to transfer the state \downdown to the state \cite{Holland2023Entanglement,Bao2023Entanglement,Picard2024Entangelement},

\begin{equation}
\ket{\Phi(T)} = -e^{-2\pi i\tilde{T}}\left[\cos(2\pi\tilde{T})\ket{\downarrow\downarrow} - i \sin(2\pi\tilde{T})\ket{\uparrow\uparrow}\right]\,,
\end{equation}
where $\tilde{T} \equiv {JT}/{4}$. This should result in spin-exchange oscillations between the states \downdown and \upup.

Figure\,\ref{fig:entanglement}(a) shows the result of applying this pulse sequence. The data points show the measured state populations; the colours are the same as in Fig.~\ref{fig:mw-spec}.
The solid lines show the expected result using our Monte Carlo model with the fitted parameters from Fig.~\ref{fig:mw-spec}.
As expected, we observe \pdowndown and \pupup oscillating in antiphase with approximate frequency $\langle J\rangle/2$.
The observed damping is caused by the non-zero value of $\sigma_J$.

We expect to prepare molecules in the maximally entangled state $\ket{\Phi^-} \equiv \left(\ket{\downarrow\downarrow} - i\ket{\uparrow\uparrow}\right)/\sqrt{2}$ (ignoring the global phase) when $T = 1/(2\langle J\rangle)$.
We find experimentally that we prepare $\ket{\Phi^-}$ with the highest probability using $T = 86(2)$\,ms, slightly faster than the $96(1)$\,ms predicted by our Monte Carlo model.
The state populations after the entanglement sequence are \pdowndown$=0.52(3)$,  \pupup$=0.47(3)$, and $P_{\uparrow\downarrow} + P_{\downarrow\uparrow} = 0.012^{+0.009}_{-0.005}$ [Fig.~\ref{fig:entanglement}(b), left panel].

We measure the fidelity $\mathcal{F}$ of entanglement by incorporating a third ``readout'' $\pi/2$ pulse after the Ramsey sequence to probe the two-particle coherence $\mathcal{C}$ \cite{Turchette1998}.
This pulse scheme is illustrated above the right panel of Fig.~\ref{fig:entanglement}(b).
The duration of the hold between the Ramsey sequence and the readout pulse is $T_h=1$\,ms and the readout pulse is performed with the same Rabi frequency as the Ramsey pulses.
We vary the phase $\phi$ of the readout pulse with respect to the Ramsey pulses.
The result is a rotation by $\pi/2$ around the axis $\cos\phi\,\hat{x} + \sin\phi\,\hat{y}$ on the Bloch sphere \cite{Holland2023Entanglement}.
From the state $\ket{\Phi^-}$, this causes oscillations in the parity $\Pi \equiv P_{\downarrow\downarrow}^{\phi}+P_{\uparrow\uparrow}^{\phi}-P_{\downarrow\uparrow}^{\phi}-P_{\uparrow\downarrow}^{\phi}$ of the form $\Pi(\phi) = \mathcal{C}\sin(2\phi)$, where $P^{\phi}$ are the state populations measured after the readout pulse \cite{Turchette1998}.

The data in right panel of Fig.~\ref{fig:entanglement}(b) show the measured behaviour of the parity $\Pi$ and the red dashed line shows a fit from which we extract $\mathcal{C} = 0.96(2)$.
The solid purple line shows the expected behaviour from our Monte Carlo model using the parameters fitted in Fig.~\ref{fig:mw-spec}.
From this model, we expect $\mathcal{C} = 0.95$, in agreement with our measured result. Our ability to perform state specific readout of both \down and \up  in a single experimental iteration allows us to eliminate most state-preparation and measurement errors through postselection \cite{Ruttley2024}.
The measured entanglement fidelity $\mathcal{F} = (P_{\downarrow\downarrow}+P_{\uparrow\uparrow} + \mathcal{C})/2 = 0.976^{+0.014}_{-0.016}$ represents the highest reported to date between a pair of molecules. 

\subsection*{Direct microwave entanglement}
Our pristine environment eliminates the need for rephasing pulses, allowing us to explore direct entanglement of molecules using microwaves. 
This development opens the door to applying quantum optimal control theory 
\cite{Glaser2015} for designing robust entangling gates between molecular rotational states \cite{Sugny2009,Pellegrini2011,Muller2011,Hughes2020}. 
Such gates are predicted to achieve fidelities of greater than $0.999$ for ultracold molecules trapped in optical tweezers \cite{Hughes2020}.

Here, we characterise the fidelity with which we can directly entangle molecules using a simple shaped pulse.
We aim to drive the transition \downdown\transitionarrow\plus whilst minimising off-resonant excitation to the state \upup. 
We choose to drive the transition with a Hann pulse of duration $\tau$ at detuning $\Delta$ during which $\Omega(t) = \Omega_0\sin^2\left(\pi t/\tau\right)$ [Fig.~\ref{fig:entanglement}(c)].
We choose optimum values of $\tau$, $\Delta$, and $\Omega_0$ by simulating the excitation with the fitted parameters from Fig.~\ref{fig:mw-spec} in order to maximise $P_{\downarrow\uparrow}+P_{\uparrow\downarrow}$ after the microwave pulse (Methods).
These (simulated) optimum values are 328\,ms, 3.069\,Hz, and 2.245\,Hz respectively.

Figure\,\ref{fig:entanglement}(c) shows measured state probabilities $P_{\downarrow\uparrow}+P_{\uparrow\downarrow}$ (empty green) and $P_{\downarrow\downarrow}+P_{\uparrow\uparrow}$ (black) as a function of $\Delta$ when $\tau$ and $\Omega_0$ are set to their optimum values.
The lines show the expected behaviour from our simulations of the system.
We predict the maximum value of $P_{\downarrow\uparrow}+P_{\uparrow\downarrow}$ is 0.96; the maximum value that we record experimentally ($0.93^{+0.04}_{-0.07}$) is within error of this prediction.

We measure the coherence $\mathcal{C}$ of the entanglement with a similar method used to characterise the entanglement generated via spin exchange.
Here, we set $\Delta = 3.3\,$Hz; the left panel of Fig.~\ref{fig:entanglement}(d) shows the populations after the Hann pulse using this detuning.
Then, for readout, we use two additional $\pi/2$ pulses ($\Omega = 882(3)\,$Hz) on the transition \down\transitionarrow\up, as illustrated above the right panel of Fig.~\ref{fig:entanglement}(d).
These pulses occur $T_h = 1$\,ms after the Hann pulse.
The first readout pulse performs the transfer $\ket{\Psi^+}\rightarrow(\ket{\uparrow\uparrow}+\ket{\downarrow\downarrow})/(\sqrt{2}i)$ and we use the second pulse to measure $\Pi$ as before.
The phase of the first readout pulse relative to the Hann pulse is arbitrary but fixed (Methods).
We vary the phase $\phi$ of the second readout pulse to obtain the oscillations in the parity $\Pi$ shown in the right panel of Fig.~\ref{fig:entanglement}(d).

From the data in Fig.~\ref{fig:entanglement}(d), we fit $\mathcal{C} = 0.93(2)$ (dashed red line).
The parity oscillation is slightly skewed towards $\Pi=1$ because molecules that are not successfully entangled preferentially occupy the state \downdown.
The measured coherence is within error of that which we expect from our simulations (0.95, solid green line).
From these measurements, we extract the entanglement fidelity $\mathcal{F} = 0.93^{+0.03}_{-0.05}$.

We emphasise that this entanglement is generated solely through microwave-driven transitions, leveraging mature microwave technology, which is easily controllable and stable. This approach enables the implementation of advanced quantum control protocols for both single- and multi-molecule gates, with potential to improve the entanglement fidelity. For instance, a pulse scheme like the one proposed by Hughes \textit{et al.} \cite{Hughes2020}, which is robust to molecules in different motional states, could mitigate dephasing in our system caused by the shot-to-shot variation in the interaction strength described by $\sigma_J$.

\subsection*{Entanglement lifetime}
To use individually trapped ultracold molecules for applications in quantum metrology \cite{DeMille2024} and quantum information processing \cite{Cornish2024}, it is highly desirable to produce long-lived entanglement.
We investigate the coherence lifetime $T_\mathcal{C}$ of entangled pairs of molecules by varying the hold time $T_h$ before applying the readout pulses.
The upper panel of Fig.~\ref{fig:entanglement}(e) shows the dependence of $\mathcal{C}$ on $T_h$ for the states $\ket{\Phi^-}$ (purple) and \plus (green).
For both states, we measure no significant change over 500\,ms. This represents a significant improvement over previous work, where entanglement coherence $T_\mathcal{C}$ was limited by single-particle coherence times and required dynamical decoupling to extend $T_2$ from 2.5(3)\,ms to 215(30)\,ms, achieving a maximum reported $T_\mathcal{C}$ of 61(3)\,ms \cite{Holland2023Entanglement}.

This long-lived entanglement paves the way for measuring sub-hertz energy shifts with quantum-enhanced metrology \cite{DeMille2024,Zhang2023DecoherenceFree}.
First, we consider the state $\ket{\Phi^-}$. 
In the rotating frame of the microwave field, a global energy difference $\Delta E$ between the states \down and \up causes $\ket{\Phi^-}$ to evolve in time $t$ to the state $\left(\ket{\downarrow\downarrow} - ie^{-i\varphi}\ket{\uparrow\uparrow}\right)/\sqrt{2}$, where $\varphi = -2\Delta Et/\hbar$. Here, the factor of 2 in the phase $\varphi$ highlights the enhanced sensitivity of this state to global perturbations, which can be leveraged to achieve Heisenberg-limited precision \cite{Huang2024}. We measure the rate of phase accumulation $d\varphi/dt$ from the measurements of the parity as a function of $T_h$, as shown in Fig.~\ref{fig:entanglement}(e) by the purple data points, and extract $2\Delta E/h = 540(90)$\,mHz. This reflects a detuning $\Delta E /h$ between the microwave field and the molecular transition frequency, allowing us to precisely measure the mean energy of the transition \down\transitionarrow\up.
In contrast, when we directly excite to the state \plus, we occupy an eigenstate of $H$ that is within a decoherence-free subspace and is immune to collective dephasing \cite{Roos2004}.
The component states (\downup and \updown) of \plus only accrue a relative phase if the energy of the transition \down\transitionarrow\up varies inhomogeneously between molecules during $T_h$, providing a sensitive probe to local perturbations.
Additionally, encoding of quantum information in these states has been demonstrated to increase the lifetime by multiple orders of magnitude \cite{Roos2004}, making these states attractive for realising quantum memories.
Any phase accrual partially transfers \plus\transitionarrow\minus, which does not couple to microwave pulses, causing the measured value of $\mathcal{C}$ to decrease whilst preserving the phase $\varphi$ of the parity oscillations.
We attribute the observed $d\varphi/dt$ for the state \plus, shown in Fig.~\ref{fig:entanglement}(e) by the green data points, to a phase drift between the microwave sources used for this measurement (Methods).
We detect no significant change in $\mathcal{C}$ over this timescale, and therefore conclude that local perturbations in the rotational splitting during $T_h$, for example from electric or magnetic field gradients, are sub-hertz.

Finally, we characterise the lifetimes of our molecules.
These are limited by Raman scattering of the tweezer light which causes apparent molecule loss due to the state specificity of our readout scheme \cite{Ruttley2024}.
Figure~\ref{fig:entanglement}(f) shows the lifetime of single molecules prepared in the states \down (blue) and \up (orange).
For this measurement, we do not postselect to remove molecule-loss errors.
For both states, the measured lifetime is 3.2(2)\,s.
Crucially, Raman scattering does not cause a bit-flip error (i.e. \down\nottransitionarrow\up and \up\nottransitionarrow\down) as it is extremely unlikely for a molecule to scatter back into the subspace $\{\ket{\uparrow},\ket{\downarrow}\}$. This represents a perfect erasure error \cite{Wu2022a} but requires detection to be effectively used for erasure error conversion \cite{Grassl1997}. This could be achieved using Rydberg atom-molecule interactions \cite{Guttridge2023} and we note that recent studies have demonstrated successful detection \cite{Liu2024a} and conversion \cite{Holland2024Erasure} of blackbody induced errors using a hybrid system of atomic and molecular ions and laser cooled molecules, respectively. 

\subsection*{Outlook}
We have realised long-lived entanglement between pairs of molecules.
Critical to this was the engineering of a pristine environment that eliminates rotational decoherence on experimental timescales.
Operating in this environment, we have prepared two-molecule Bell states using dipolar spin exchange and direct microwave excitation with fidelities $0.976^{+0.014}_{-0.016}$ and $0.93^{+0.03}_{-0.05}$ respectively.
This represents the highest reported entanglement fidelity for individually trapped polar molecules to date and the first realisation of a two-molecule microwave gate.
Further, these methods prepare Bell states that are sensitive to either the global or local environment, realising sensitive probes of different physical phenomena.

In the near term, the speed and fidelity of our Bell-state preparation may be improved by changing the confinement of the molecules to access smaller separations. For example, transferring the molecules into a magic-wavelength optical lattice should give access to sub-micrometre separations and increased molecular confinement, resulting in increased interaction strengths with reduced noise. 
Such improvements will allow the implementation of high-fidelity two-molecule gates \cite{Ni2018,Hughes2020} that entangle molecules on the millisecond timescale, whilst preserving the pristine environment and long-lived entanglement associated with magic-wavelength trapping.

Further ahead, our results show that there are no fundamental obstacles to using ultracold molecules for a wide range of applications in quantum science. The ability to prepare molecules in various Bell states opens up new avenues for studying quantum interference effects in ultracold chemistry \cite{Liu2024}. Furthermore, the deterministic preparation of molecules in a decoherence-free subspace paves the way for quantum-enhanced metrology \cite{Zhang2023DecoherenceFree} and the use of long-lived rotational states as quantum memories within hybrid quantum systems \cite{Zhang2022,Wang2022,Guttridge2023}. 
Finally, our modelling suggests that second-scale coherence will be simultaneously achievable for multiple rotational transitions \cite{Guan2021,Gregory2024}; this will allow the ladder of molecular rotational states to be exploited as qudits \cite{Sawant2020} or synthetic dimensions \cite{Sundar2018}.  


\clearpage
\newpage
\section*{Methods}
\small 

\noindent\textbf{Experimental apparatus} \\
In our experimental apparatus~\cite{Brooks21,Spence22}, we produce ultracold RbCs molecules trapped in one-dimensional arrays of optical tweezers at wavelength 1065.512\,nm (hereafter 1066\,nm).
The molecules are trapped inside an ultra-high vacuum glass cell, with the tweezers formed by focusing light through a high numerical aperture objective lens prior to this cell. The molecules are formed by associating Rb and Cs atoms as described in Ref.~\cite{Ruttley2024}.
\\

\noindent{\it Magic-wavelength tweezers.} 
For the work presented here, we have added a set of tweezers at a magic wavelength of 1145.31\,nm which eliminates the differential ac Stark shift for the rotational transition \down\transitionarrow\up.
Prior to the objective lens, the polarisation of the tweezers is parallel to the quantisation axis set by the external magnetic field.
The array of tweezers is created with an acousto-optic modulator (AOM) prior to the objective lens [Extended Data Fig.~\ref{fig:optical_setup}(a)].
By applying multiple radio-frequency (RF) tones to the AOM, we form multiple diffracted beams to generate the tweezers. 
We dynamically switch and move the tweezers by changing the RF tones applied to the AOM to manipulate the trapped molecules mid-routine.
By imaging Cs atoms trapped in the magic tweezers, we calibrate the change in tweezer position (at the focal plane) with the change in RF frequency applied to the AOM as $397(7)\,\mathrm{nm}/\mathrm{MHz}$.

We perform parametric heating measurements \cite{Savard1997} of Cs atoms trapped in the magic tweezers to characterise their $1/e^2$ beam waists.
To do this, we modulate the intensity of the traps and measure a loss feature that occurs when the modulation frequency is twice that of the trap frequency.
We assume the light in the focal plane is well described by a Gaussian beam and take the polarisability of the Cs atoms to be $1160(1)\times4\pi\varepsilon_0 a_0^3$ \cite{UDportal} to obtain the $1/e^2$ waist $1.87(5)\,\mu$m. 

For efficient transfer of molecules between different tweezer arrays, it is important that they are well overlapped.
We overlap the tweezers in the radial directions by imaging Cs atoms in both sets of tweezers, and moving the magic tweezers until the positions of the atoms overlap.
This allows us to overlap the centre of the tweezers to sub-micrometre accuracy.
This method is much less sensitive to the overlap in the direction of tweezer-light propagation. 
We coarsely overlap the arrays in this direction by moving a lens in the expansion telescope of the 1145\,nm light so that atoms in both arrays are in focus on our imaging camera.
We expect that there could be an alignment error of up to few micrometres in this direction.

To transfer molecules between the two arrays, we start with the tweezers overlapped.
We ramp up the power of the magic tweezers and then ramp down the power of the 1066\,nm array.
During this step, the separation between neighbouring tweezers is approximately $4\,\mu$m.
After this transfer, we switch off excess tweezers to discard excess molecules.
At the end of an experiment, we transfer molecules back to the 1066\,nm array before disassociating them and reimaging their constituent atoms.
During this process, we map the internal state of the molecule onto atomic position for multistate readout \cite{Ruttley2024}.

To tune the dipolar interaction strength between molecules, we tune the separation of the molecules by chirping the frequency of the RF tones that generate their tweezers.
For all the experiments presented in the main text, we move a pair of molecules symmetrically around their mean position to minimise the chance that one molecule is preferentially heated during the movement process.
\\

\noindent{\it Magic-frequency stabilisation.} 
In previous work trapping RbCs molecules in magic-wavelength traps \cite{Gregory2024}, the single-molecule coherence time was limited by the frequency stability of the trapping laser.
The laser was stabilised to a cavity of finesse $\sim 400$, and a frequency stability (standard deviation) of $0.76$\,MHz was achieved.

For this work, we reference an external-cavity diode laser (ECDL, Toptica DL pro) at 1145\,nm to an ultra low expansion cavity (Stable Laser Systems) with a finesse of $\sim3.7\times10^{4}$.
We stabilise this laser with a fast feedback loop (Toptica FALC pro) and achieve a linewidth of $\sim 5$\,kHz.
To allow for future scaling to larger arrays, we source the trapping light from a vertical-external-cavity surface-emitting laser (Vexlum VALO) which provides up to 4\,W of optical power.
We stabilise the beat note between this laser and the ECDL. Feedback to the laser frequency is achieved using a piezo-electric actuator mounted to a mirror in the laser cavity.
With stabilisation, the standard deviation of the beat-note signal is 80(20)\,kHz.
Therefore, we expect the frequency of the trapping light to be stable to within 80(20)\,kHz.
\\

\noindent{\it Tweezer-intensity noise.}
In Fig.~\ref{fig:ramsey-magic}(c), we show the measured single-molecule coherence times $T_2^*$ as a function of the detuning $\Delta_\mathrm{magic}$ of the tweezers from the magic frequency $f_\mathrm{magic}$.
We model the effect of intensity noise in our experiment to understand the behaviour of $T_2^*$ with $\Delta_\mathrm{magic}$ and briefly discuss that model here.

We determine $f_\mathrm{magic}$ and the sensitivity of the molecules to $\Delta_\mathrm{magic}$ with a Ramsey procedure \cite{Hepworth2024}.
The differential ac Stark shift $h\Delta\alpha_\mathrm{ac}$ is proportional to the power $P$ of each tweezer and $\Delta_\mathrm{magic}$.
The scaling constant $k =  923(3)\,\mathrm{mHz}/\mathrm{MHz}/\mathrm{mW}$ relates these such that $\Delta\alpha_\mathrm{ac} = k \Delta_\mathrm{magic} P$.
The power in each tweezer is measured prior to the objective lens; we estimate that the transmission from this location to the science cell is $0.48(1)$.

To model the intensity noise, we assume that there is Gaussian noise on $P$ such that it is sampled from a Gaussian distribution with mean $\langle P \rangle$ and standard deviation $\sigma_P$.
For the measurement in Fig.~\ref{fig:ramsey-magic}, $\langle P\rangle = 0.36\,$mW.
This noise is mapped to $\Delta\alpha_\mathrm{ac}$ with standard deviation $\sigma_\alpha = k \Delta_\mathrm{magic} \sigma_P$. 
Therefore, the Ramsey contrast $C(T) = \exp\left[-(2\pi\sigma_\alpha T)^2/2\right] \equiv \exp\left[-(T/T_2^*)^2\right]$.
Hence, $T_2^* = 1/\left(\sqrt{2}\pi\sigma_\alpha\right)$ and the solid line in Fig.~\ref{fig:ramsey-magic}(c) shows the predicted behaviour when $\sigma_P/\langle P \rangle = 0.7\,\%$. 
\\

\noindent{\it Achieving magic trapping conditions for multiple tweezers.}
For the experiment in Fig.~\ref{fig:ramsey-magic}, we prepare molecule pairs in tweezers at a separation of $8.6(2)\,\mu$m.
They are generated using a frequency difference of $\Delta f = 21.7\,$MHz between the two RF tones applied to the AOM and the power per tweezer is actively stabilised to $\langle P \rangle = 0.36$\,mW.
Therefore, we expect $\Delta\alpha_\mathrm{ac}$ would be different by $\delta = k\langle P\rangle\Delta f = 7.2$\,Hz.
The data in Fig.~\ref{fig:ramsey-magic}(a) are fitted with a damped sinusoidal function with frequency $\nu$.
For the tweezer that is closer to $f_\mathrm{magic}$ (blue filled points), we fit $\nu = 999.26(2)$\,Hz, and for the tweezer that is further detuned (red empty points), we fit 
$\nu = 992.49(1)$\,Hz. 
This is a frequency difference of 6.77(3)\,Hz, approximately 6\% smaller than expected. 

For the experiments in Figs~\ref{fig:mw-spec} and \ref{fig:entanglement}, we prepare molecule pairs in tweezers at a separation of $2.78(5)\,\mu$m.
Each tweezer has a time-averaged power of $\langle P \rangle \sim 0.3$\,mW and is generated by RF tones with a frequency difference of $\Delta f = 7.011$\,MHz.
This difference in detuning from the magic frequency, results in a difference in transition frequency between the two molecules of $\delta = k\langle P\rangle\Delta f \sim 2$\,Hz.

In order to engineer the regime $\delta\ll J$, we minimise $\delta$ by minimising $\Delta f$ whilst maintaining the same tweezer separation. 
To do this, we modulate the tweezer intensities in antiphase at a frequency of $500$\,kHz with a duty cycle of $0.35$.
Simultaneously, we modulate the frequency of an RF tone applied to a compensation AOM so that, ideally, the light forming the two tweezers has identical frequency.
A schematic of the modulation scheme is shown in Extended Data Fig.~\ref{fig:optical_setup}(b).
The $500$\,kHz modulation frequency is far above any parametric resonances and we do not observe any change in the molecule loss rate due to the modulation.
We do not actively stabilise the tweezer intensity when operating in this regime.
We have verified that this modulation does not affect single-molecule coherence by repeating measurements such as those in the upper panel of Fig.~\ref{fig:mw-spec}(c) with and without this modulation.
We attribute the non-zero value of $\delta$ reported in the main text to the non-zero decay time of tones in the amplifier that drives this compensation AOM.

In future, we plan to scale to larger molecule arrays with methods that will not require this compensation AOM.
For example, by using a spatial light modulator to form the magic tweezers, as we do for the 1066\,nm tweezers \cite{Ruttley2024}, all tweezers will have the same frequency.
Alternatively, a pair of crossed acousto-optic deflectors could be used to create arrays of magic-wavelength tweezers with a constant frequency across the array \cite{Barnes2022}. Additionally, we note that all sites in a magic-wavelength (1D) optical lattice would have the same frequency.
\\

\noindent{\it Microwave excitation.}
In our experiment, we prepare RbCs molecules in the absolute internal ground state $\ket{\downarrow} = \ket{N=0,M_N=0,m_\mathrm{Rb}=3/2,m_\mathrm{Cs}=7/2}$.
Here, $N$ is the rotational quantum number, $M_N$ is its projection, and $m_{\mathrm{Rb}}$ ($m_{\mathrm{Cs}}$) is the projection of the nuclear spin of Rb (Cs).
We couple this state to the excited rotational state $\ket{\uparrow} = \ket{N=1,M_N=1,m_\mathrm{Rb}=3/2,m_\mathrm{Cs}=7/2}$. 
Both of these states are stretched with maximum projections of angular momentum.
In our experiment, the quantisation axis is set by the externally applied magnetic field ($\sim181.7$\,G) which stays approximately constant for all science stages of the experiment.

The transition \down\transitionarrow\up is magneticially insensitive. 
The dominant contribution to the Zeeman shifts of the states \down and \up is associated with the projection of the nuclear spins. 
However, as these are both stretched states with the same $m_{\mathrm{Rb}}$ and $m_{\mathrm{Cs}}$, their nuclear-spin Zeeman shifts are equal.
The rotational Zeeman effect is very small \cite{Aldegunde2008,Gregory2016}, leading to a differential Zeeman shift of $\sim 5\,\mathrm{Hz}/\mathrm{G}\times h$. 
In our experiment, we stabilise the magnetic field to the $\sim 10$\,mHz level so that the differential shift does not vary shot-to-shot.

We drive the molecular transition \down\transitionarrow\up with microwaves radiated from a dipole Wi-Fi antenna mounted approximately 10\,cm from the vacuum chamber.
The frequency of the transition in free space (or in a perfectly magic tweezer) is $980.38559837(4)$\,MHz \cite{Hepworth2024}.
The resultant microwaves are not well polarised, so it would be possible to drive transitions to other rotational states.
For this reason, we use Rabi frequencies $\lesssim 10$\,kHz such that off-resonant excitation to other states is negligible \cite{Ruttley2024} and each molecule can be considered a two-level system. 
For kilohertz-scale Rabi frequencies, we drive the antenna with an Agilent E4400B source and typically input a microwave power of $\sim 0$\,dBm to the antenna.
We vary the phase of this source when measuring the parity $\Pi$ presented in Fig.~\ref{fig:entanglement}.
For hertz-scale Rabi frequencies, we use an Anritsu MG369xC source set to $\sim-15$\,dBm with a further $55\,$dB of attenuation. 
We amplitude modulate this source with an arbitrary function generator (Tektronix AFG3022C) when using the Hann pulse for direct microwave entanglement.
The sources are combined before the antenna with an RF switch (Minicircuits ZFSWA2R-63DR+) and are referenced to the same $10\,$MHz GPS signal to maintain a constant, but arbitrary, relative phase.
\\

\noindent\textbf{Experimental statistics} \\
To obtain statistics, we repeat each experimental sequence many times. 
Data points in figures show the average state populations from these repeats and error bars show the $1\sigma$ binomial confidence intervals, calculated using the Jeffreys prior \cite{Jeffreys1946Interval,Brown2001Intervals,Cai2005Intervals}, and are indicative of the number of repeats used to obtain each data point.
Most data presented here are obtained by postselecting to ignore experimental runs in which molecule formation was unsuccessful or molecules were not detected in the states \down or \up \cite{Ruttley2024}.
The exception for this is the data presented in Fig.~\ref{fig:entanglement}(f) where we measure the molecule lifetimes and so only postselect to remove detectable molecule-formation errors.

With postselection, we can obtain statistics for single- and two-molecule cases in a single set of experimental runs using the same sequence.
For example, for each value of $\Delta$ in Fig.~\ref{fig:mw-spec}(b), we repeat the experiment $\sim400$ times.
In 25\% of runs we successfully form and detect exactly one molecule in either the state \down or \up.
Therefore, each data point in the upper panel represents $\sim 100$ samples of the binomial distribution, and the error bars are calculated accordingly.
Likewise, in 7\% of runs we successfully form and detect exactly two molecules, and each data point in the lower panel reflects $\sim 30$ samples.
\\

\noindent\textbf{Expected interaction strength} \\
Here, we consider the strength of the spin-exchange interaction between the molecular pair states \downup and \updown.
First, we consider the case where the molecules can be treated as point particles with zero temperature.
Then, we estimate the effect that the non-zero temperature and wavefunction size has on this interaction strength.
\\


\noindent{\it Point-particle and zero-temperature case.} 
The strength of the dipole-dipole interaction between the states \downup and \updown is \cite{Wall2015}
\begin{equation}
    J = -\frac{1}{h}\frac{1-3\cos^2(\theta)}{|\mathbf{r_1} - \mathbf{r_2}|^3} \frac{d_{\downarrow\uparrow}^2}{4\pi\varepsilon_0}\,.
\end{equation}
Here, $\mathbf{r_i}$ is the position vector of molecule $i$ and $\theta$ is the angle between the quantisation axis and the intermolecular vector.
$\varepsilon_0$ is the vacuum permittivity. 
$d_{\downarrow\uparrow}\equiv\braket{\uparrow|\hat{d_1}|\downarrow}$ is the relevant matrix element for the dipole operator $\hat{d_1}$ that corresponds to the $\sigma^+$ transition that we use.
At zero electric field, $d_{\downarrow\uparrow} = d/\sqrt{3}$, where $d = 1.225(11)$\,D is the RbCs molecule-frame electric dipole moment \cite{Molony2014}.

For all experiments here, the intermolecular axis is parallel to the quantisation axis (i.e.\ $\theta = 0$).
We apply no external electric field, and assume that the stray electric field is negligible.
For the experiment presented in Fig.~\ref{fig:ramsey-magic}, we prepare molecules at a separation $|\mathbf{r_1} - \mathbf{r_2}| = 8.6(2)\,\mu$m.
Therefore, if the molecules were point particles pinned to the centre of their respective optical tweezer, we would expect $J = 0.24(1)$\,Hz, where the uncertainty reflects the uncertainty on the molecular separation.
Likewise, for the experiments in presented Figs~\ref{fig:mw-spec} and \ref{fig:entanglement}, $|\mathbf{r_1} - \mathbf{r_2}| = 2.78(5)\,\mu$m giving $J = 7.0(4)$\,Hz.
In both cases, the uncertainty in $J$ reflects the uncertainty on the molecular separation.
\\


\noindent{\it Effect of motional excitation.} 
We fit the microwave spectroscopy shown in Fig.~\ref{fig:mw-spec} with a Monte Carlo model where $J$ is sampled from a normal distribution for every iteration of the experiment.
Using this model we extract the mean $\langle J\rangle= 5.20(5)$\,Hz and standard deviation $\sigma_J = 1.0(1)$\,Hz.

We expect that motional excitation of the molecules causes the reduction in $\langle J\rangle$ from the expected value and is the dominant contribution to $\sigma_J$.
To estimate the magnitude of this effect, we numerically calculate the matrix elements
\begin{equation}
\bar{J}(\boldsymbol{n_1};\boldsymbol{n_2}) = -\frac{1}{h} \frac{d_{\downarrow\uparrow}^2}{4\pi\varepsilon_0}\Braket{\boldsymbol{n_1}\boldsymbol{n_2}|\frac{1-3\cos^2(\theta)}{\left|\mathbf{r_1} - \mathbf{r_2}\right|^3}|\boldsymbol{n_1}\boldsymbol{n_2}} \,,
\end{equation}
where $\ket{\boldsymbol{n_i}} \equiv \ket{n^i_x,n^i_y,n^i_z}$ is the three-dimensional wavefunction for molecule $i$ which is labelled by the number of motional quanta in each of the three directions. 
Here, we define the $x$-axis as the quantisation axis, the $y$-axis as the other radial axis of the tweezers, and the $z$-axis as the direction of tweezer-light propagation, as shown in Extended Data Fig.~\ref{fig:optical_setup}(a).
We assume that the trapping potential is harmonic and the three axes are separable such that
\begin{equation}
    \braket{\boldsymbol{r_i}|\boldsymbol{n_i}} = \prod_{r\in\{x_i,y_i,z_i\}} C(n_r) H_{n_r}(r/\beta_r) e^{-r^2/2\beta^2}\,,
\end{equation}
where $H_{n_r}$ are the Hermite polynomials and the index $r$ runs over the three separable axes. $\beta_r = 2\pi\sqrt{{m\nu_r}/{h}}$ and $\nu_r$ are the confinement length and trap frequency along the $r$-axis respectively, and the normalisation constant  $C(n_r) = 1/\sqrt{(2^{n_r}{n_r}!\beta_r\pi^{1/2})}$.



Extended Data Fig.~\ref{fig:motional_excitation} shows calculations of selected values of $\bar{J}$. 
In general, $\bar{J}$ is a six-dimensional matrix; we show the three slices of this matrix where the motional quanta of the molecules along one axis is varied while the is no motional excitation along the other axes.
For this calculation, the separation between the most-likely positions of the molecules is $2.78\,\mu$m along the $x$-axis.
The molecules are trapped in tweezers of waist $1.87\,\mu$m and intensity $4$\,\kwcmsq.
We neglect the effect of the tweezer confining the first molecule on the second molecule (and vice versa) and assume that fluctuations in the relative positions of the tweezers are negligible as they are formed from a common source \cite{Guttridge2023}.
We take the polarisability of the molecules at the magic wavelength to be $360\times4\pi\varepsilon_0 a_0^3$ 
\cite{Guan2021} such that the trap frequencies are $\nu_x = \nu_y = 1.9$\,kHz and $\nu_z = 0.3$\,kHz.

We estimate that 58(6)\% of molecules formed in the 1066\,nm array occupy the three-dimensional motional ground state \cite{Zhang2020,Ruttley2024}. 
Further, we expect that the vast majority of the motionally-excited molecules have just one motional quantum.
Therefore, the most likely scenario is that, when a pair of molecules is formed, one occupies the motional ground state and the other has one motional quantum.
Assuming negligible heating as the molecules are transferred to the magic tweezers, the relevant matrix element $\bar{J}\approx5.4(3)\,$Hz.
This is approximately equal to our measured value of $\langle J\rangle$, and the stochastic occupancy of the motional states will give rise to $\sigma_J$.

In future, we expect that moving to more confining traps (e.g.\ by trapping the molecules in an optical lattice) will allow smaller separations and reduce the wavefunction spread, leading to an increase in $\langle J\rangle$ and a reduction in $\sigma_J$. 
We note that $\sigma_J$ could also be reduced by increasing the fraction of molecules that occupy the three-dimensional motional ground state by reducing atomic heating prior to association \cite{Spence22,Zhang2020}.
\\


\noindent\textbf{Simulations of $H$} \\
To simulate the dynamics of $H$, we use the Python package QuTiP \cite{Johansson2012} and model the time evolution of the two-molecule system with different microwave pulses and hold times.\\

\noindent{\it Eigenstates in the absence of microwaves.} 
Equation~\eqref{eq:ham} gives the Hamiltonian $H$ that describes our system of two interacting molecules.
In the absence of microwave radiation, $H$ simplifies to
\begin{equation}
    H_0 = \frac{h}{2} \begin{pmatrix}
        0 & 0 & 0 & 0\\
        0 & \delta & J & 0\\
        0 & J & -\delta & 0\\
        0 & 0 & 0 & 0
    \end{pmatrix}\,,
\end{equation}
in the basis $\{\ket{\downarrow\downarrow},\ket{\downarrow\uparrow},\ket{\uparrow\downarrow},\ket{\uparrow\uparrow}\}$.
The eigenstates of $H_0$ are: \downdown, \upup, 
\begin{align}
    \ket{\tilde{\Psi}^+} &= N_+\begin{pmatrix}
        0 \\
        J \\
        \sqrt{J^2+\delta^2} - \delta \\
        0 
    \end{pmatrix}\,\mathrm{, and}\\
    \ket{\tilde{\Psi}^-} &= N_-\begin{pmatrix}
    0 \\
    \sqrt{J^2+\delta^2} - \delta \\
    - J \\
    0 
    \end{pmatrix}\,,
\end{align}
where $N_\pm$ are normalisation constants.

In the main text, we consider the limit of strong interactions (i.e.\ $|J|/|\delta|\rightarrow\infty$) where $\ket{\tilde{\Psi}^\pm}\rightarrow\ket{\Psi^\pm} \equiv (\ket{\downarrow\uparrow}\pm\ket{\uparrow\downarrow})\sqrt{2}$.
However, the non-zero value of $\delta$ in our experiment gives rise to eigenstates that are slightly asymmetric.
The eigenstates for our system, taking $J = 5.20\,$Hz and $\delta = 220\,$mHz, are $\ket{\tilde{\Psi}^+} = 0.722\ket{\downarrow\uparrow} + 0.692\ket{\uparrow\downarrow}$ and $\ket{\tilde{\Psi}^-} = 0.692\ket{\downarrow\uparrow} - 0.722\ket{\uparrow\downarrow}$, where the coefficients are given to three significant figures.

In the lower panel of Fig.~\ref{fig:mw-spec}(b), we show microwave spectroscopy in which we drive the transition $\ket{\downarrow\downarrow}\rightarrow\ket{\tilde{\Psi}^+}$.
The asymmetry in the probability amplitudes \downup and \updown in $\ket{\tilde{\Psi}^+}$ is the reason why we measure slightly higher population in the state \downup than in the state \updown. 
This has only a slight effect on the achieved entanglement fidelity, the dominant limitation to which is the non-zero value of $\sigma_J$.
\\

\noindent{\it Design of direct-entanglement pulse.}
For the demonstration of the two-molecule microwave gate shown in Fig.~\ref{fig:entanglement}(c), we use a simple shaped pulse.
We choose the parameters of this pulse using our Monte Carlo model with the parameters fitted from the data in Fig.~\ref{fig:mw-spec}.

First, we model and optimise the pulse assuming that there is no noise in $J$. 
We consider three simple pulse shapes: a square pulse ($\Omega(t) = \Omega_0$ for $0<t<\tau$, $0$ otherwise), a Hann pulse ($\Omega(t) = \Omega_0\sin^2\left(\pi t/\tau\right)$), and a Blackman-Harris pulse ($\Omega(t) = \Omega_0\left[a_0 - a_1 \cos\left(2\pi t/\tau\right) + a_2 \cos\left(4\pi t/\tau\right) - a_3 \cos\left(6\pi t/\tau\right)\right]$ for $a_0=0.35875$, $a_1=0.48829$, $a_2=0.14128$, and $a_3=0.01168$).
Here, $\Omega(t)$ is the Rabi frequency which we drive the single-molecule transition \down\transitionarrow\up, $\Omega_0$ is the peak Rabi frequency, and $\tau$ is the pulse duration.
For each pulse shape, we vary $\Omega_0$ and calculate $P_{\downarrow\uparrow}+P_{\uparrow\downarrow}$ as a function of $\tau$ and the microwave detuning $\Delta$ [e.g.\ Extended Data Fig.~\ref{fig:pulse_design}, inset].
$P_{\downarrow\uparrow}+P_{\uparrow\downarrow}$ is a good proxy for the fidelity of the entangling gate, because pairs that are not entangled preferentially occupy the states \downdown and \upup.
This gives an optimum value of $\tau$ and $\Delta$ for each value of $\Omega_0$, with an associated maximum $(P_{\downarrow\uparrow}+P_{\uparrow\downarrow})_\mathrm{max}$.
We show the behaviour of $(P_{\downarrow\uparrow}+P_{\uparrow\downarrow})_\mathrm{max}$ on $\tau$ in the top panel of Extended Data Fig.~\ref{fig:pulse_design}; a longer pulse duration generally allows higher fidelity entanglement because a smaller Rabi frequency can be used to minimise off resonant excitation to \upup.

We now consider fluctuations in $J$.
With the optimum pulse parameters obtained above, 
we use our Monte Carlo model to recalculate $(P_{\downarrow\uparrow}+P_{\uparrow\downarrow})_\mathrm{max}$ when $\sigma_J = 1\,$Hz [Extended Data Fig.~\ref{fig:pulse_design}, bottom panel].
The effect of $\sigma_J$ is to favour larger Rabi frequencies (i.e. smaller $\tau$) which spectrally broaden the excitation feature.
We expect that, out of the pulse shapes considered, a Hann pulse will achieve the highest $(P_{\downarrow\uparrow}+P_{\uparrow\downarrow})_\mathrm{max}$.
The corresponding pulse parameters are $\tau = 328$\,ms, $\Delta = 3.069$\,Hz, and $\Omega_0 = 2.245$\,Hz and we use these for the experiments presented in Fig.~\ref{fig:entanglement}. 
\\

\noindent\textbf{Data availability} \\
The data that support the findings of this study are available
at \url{https://doi.org/10.15128/r1bv73c047f}.
\\

\noindent\textbf{Acknowledgments} \\
We thank Fritz von Gierke for assistance with the installation of the magic-wavelength tweezers. We acknowledge support from the UK Engineering and Physical Sciences Research Council (EPSRC) Grants EP/P01058X/1, EP/V047302/1, and EP/W00299X/1, UK Research and Innovation (UKRI) Frontier Research Grant EP/X023354/1, the Royal Society, and Durham University. 
\\

\noindent\textbf{Author contributions} \\
\textbf{D.K.R.} Investigation (equal), Methodology (equal), Formal analysis (lead), Visualisation (lead), Software (supporting), Writing -- original draft, Writing -- Review \& Editing (equal), Data Curation.\\
\textbf{T.R.H.} Investigation (equal), Methodology (equal), Formal analysis (supporting), Visualisation (supporting), Software (lead), Writing -- Review \& Editing (equal).\\
\textbf{A.G.} Investigation (supporting), Methodology (equal), Conceptualisation (equal), Writing -- Review \& Editing (equal).\\
\textbf{S.L.C.} Methodology (equal), Conceptualisation (equal), Supervision, Funding acquisition, Writing -- Review \& Editing (equal).
\\

\noindent\textbf{Competing interests} \\
The authors declare no competing interests.
\\

\noindent\textbf{Correspondence and requests for materials} should be addressed to Simon L. Cornish.

\setcounter{figure}{0}
\newcounter{EDfig}
\renewcommand{\figurename}{Extended Data Fig.}

\clearpage
\newpage

\begin{figure*}
    \centering
    \includegraphics[width=\columnwidth]{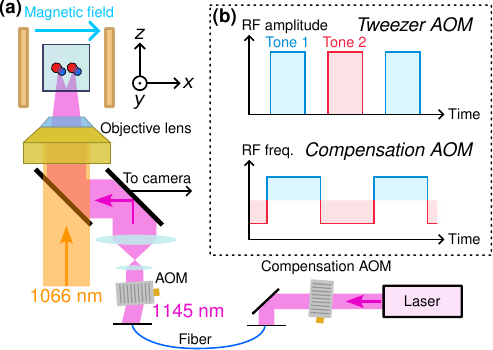}
    \caption{\textbf{Generation of multiple magic-wavelength tweezers.} (a) Simplified optical setup of the magic tweezers used in our experiment. We generate an 1145\,nm tweezer array by driving an AOM with multiple RF tones. An additional compensation AOM can be used to modulate the frequency of the input light. (b) Schematic of the modulation scheme used to generate two time-averaged tweezer traps with the same laser frequency. Upper: the RF amplitudes of the two frequency tones used to drive the tweezer AOM in order to generate two time-averaged traps. Lower: simultaneous switching of the RF frequency with which we drive the compensation AOM ensures that the light delivered to the molecules has the same frequency for both tweezers.}
    \refstepcounter{EDfig}\label{fig:optical_setup}
\end{figure*}

\begin{figure*}[t]
    \centering
    \includegraphics[width=\textwidth]{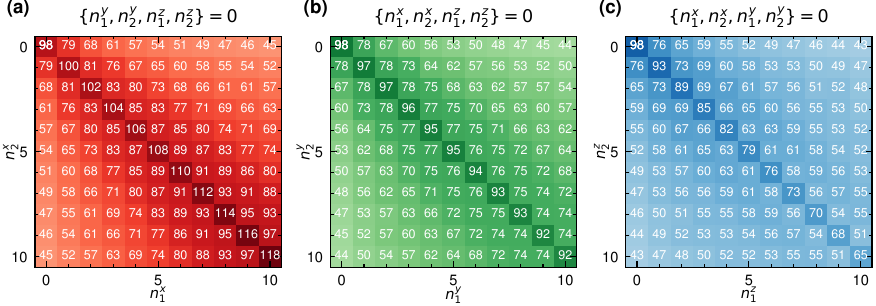}
    \caption{\textbf{Calculated  matrix elements $\bar{J}$.} Matrix elements for molecules motionally excited along the (a) $x$- (b) $y$- and (c) $z$-axes. The modal intermolecular separation is $2.78\,\mu$m along the $x$-axis. The colours and numbers label the value of $\bar{J}/J$ (in percent) where $J = 7.0$\,Hz is the expected value for a point-particle at zero-temperature.}
    \refstepcounter{EDfig}\label{fig:motional_excitation}
\end{figure*}

\begin{figure*}[t]
    \centering
    \includegraphics[width=\columnwidth]{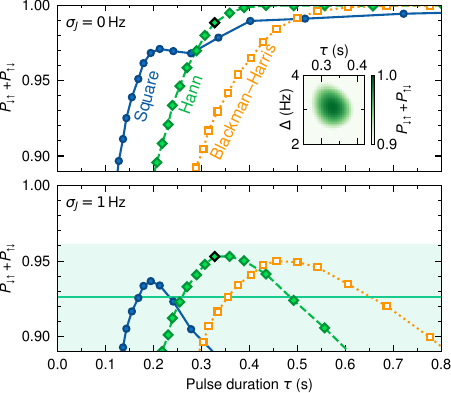}
    \caption{\textbf{Optimisation of a microwave pulse for direct entanglement.} Top: Predicted state populations $P_{\downarrow\uparrow}+P_{\uparrow\downarrow}$ for different pulse durations $\tau$ when there is no noise in $J$. The peak Rabi frequency $\Omega_0$ is set for each value of $\tau$ to achieve the maximum value of $P_{\downarrow\uparrow}+P_{\uparrow\downarrow}$. The inset shows $P_{\downarrow\uparrow}+P_{\uparrow\downarrow}$ as a function of $\Delta$ and $\tau$ when $\Omega_0 = 2.245\,$Hz and a Hann pulse is used. The peak value $(P_{\downarrow\uparrow}+P_{\uparrow\downarrow})_\mathrm{max}$ and the optimum value of $\tau$ correspond to the highlighted point in the main figure. This is simulated for various $\Omega_0$ to find the optimum parameters. Bottom: As above, but when $\sigma_J = 1$\,Hz. The highlighted point corresponds to the pulse parameters used when taking the data presented in Fig.~\ref{fig:entanglement}. The horizontal line shows the peak value of $P_{\downarrow\uparrow}+P_{\uparrow\downarrow}$ measured in Fig.~\ref{fig:entanglement}(c) and the shaded region shows the $1\sigma$ confidence interval.}
    \refstepcounter{EDfig}\label{fig:pulse_design}
\end{figure*}

\end{document}